\documentclass[pre,twocolumn,showpacs,preprintnumbers]{revtex4-1}
\usepackage{amsmath,amssymb,latexsym} 
\usepackage{color}

\usepackage{graphicx} 
\usepackage{subfigure}
\usepackage{natbib}
\usepackage{epstopdf}
\usepackage{color}
\usepackage{makecell}
\usepackage[table]{xcolor}
\usepackage{tabularx}
\usepackage{ragged2e}

\begin{document}

\title
{Reply to comment by  Witte {\it et al.}  on \\ 
 ``Isochoric, isobaric, and ultrafast conductivities of\\
 aluminum, lithium,and carbon in the warm dense matter regime'', \\   
 Phys. Rev. E {\bf 96}, 053206 (2017).
}

\author
{M.W.C. Dharma-wardana}\email[Email address:\ ]
{chandre.dharma-wardana@nrc-cnrc.gc.ca} 
\author
{D. D. Klug}
\affiliation{
National Research Council of Canada, Ottawa, Canada, K1A 0R6
}

\author{L. Harbour and Laurent J. Lewis}
\affiliation{
D\'{e}partement de Physique, Universit\'{e}
 de Montr\'{e}al, Montr\'{e}al, Qu\'{e}bec, Canada.} 

%
\date{\today}

\begin{abstract}
In Phys. Rev. E, {\bf 99}, 047201 (2019) Witte {\it et al.} have commented
 on our    conductivity calculations
[Phys. Rev. E {\bf 96}, 053206 (2017)] for warm dense matter (WDM).
(i) They criticize our use of the spherically-averaged structure factor $S(k)$
 for calculations of the static conductivity $\sigma$
of FCC aluminum - a common approximation for polycrystalline
materials. They themselves give no calculations
as their  method using  density-functional theory (DFT)
 and molecular dynamics (MD) based  Kubo-Greenwood (KG)  calculations
 becomes impractical for cold ions.
(ii) We are satisfied that Witte {\it et al.}
no longer claim  a factor of $\sim$ 1.5 change in $\sigma$
 on changing the exchange-correlation (XC) functional used.
(iii) They have provided computer-intensive  
calculations of $\sigma$ for aluminum using DFT-MD-KG
 simulations,  for temperatures $T$ up to 15 eV but using only $N$=64 atoms
 in the simulation, where as a mixture of ionic species needs a far larger $N$
to be credible. We present  multi-species  conductivity calculations
via a parameter-free DFT theory  [Phys. Rev. E.
{\bf 52}, 5352 (1995)] for 5 eV to  50 eV.
(iv)
The  conductivities   obtained from
well-converged  DFT-MD-KG methods
show a significant  underestimate of $\sigma$; this is especially evident
for  the isochoric conductivity  $\sigma_{\rm ic}$  extrapolating
to $\sim3.5\times 10^6$ S/m, i.e, {\it even below} the experimental
{\it isobaric} value  of 4.1$\times 10^6$ S/m  at
the melting point, when a value
of $\sim 5\times 10^6$ S/m is anticipated.
\end{abstract}
\pacs{52.25.Os,52.35.Fp,52.50.Jm,78.70.Ck}
\maketitle

\section{Introduction}
\label{intro} 
Witte {\it et al.}\cite{WitteComment19}  
have  commented on our work,
Ref.~\cite{cdw-plasmon16}, Ref.~\cite{cond3-17},
referred to here as DW16 and  DKHL respectively, while Witte {\it et al.}
refer to both as DWD.
Their comment is denoted here as ``WitteC ''. As DW16
involves only one author, this reply is  confined
to comments on DKHL, while DW16-issues  will be addressed
elsewhere. 
 The static conductivities $\sigma$ of 
Al, C and  Li under warm dense matter (WDM) conditions were studied
 in DKHL using
the {\it simplest implementation}
of the neutral-pseudo-atom (NPA) model, as described in DKHL.
 
\begin{figure}[t]
\includegraphics[width=\columnwidth]{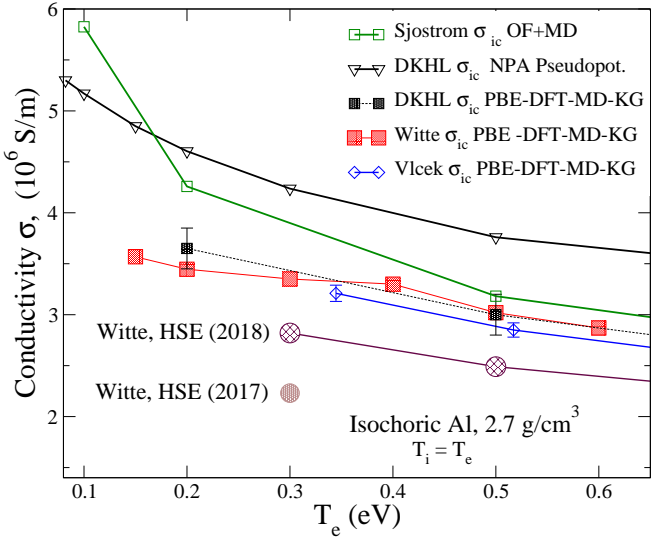}
\caption{(Color online) Isochoric conductivities $\sigma_{\rm ic}$
 of aluminum from near
 its melting point to about 0.6 eV; 
our DFT+MD data and those of Vl\v{c}ek \cite{Vlcek12}, Sjostrom
 {\it et al.} \cite{SjosCond15} are shown.
The Witte {\it et al.}~\cite{WittePRL17} calculation of $\sigma_{\rm ic}$
at $T=0.3$ eV and $\rho=2.7$ g/cm$^3$ using the PBE and
HSE functionals in 2017, and in 2018,
are also shown. 
}
\label{replyXC.fig}
\end{figure}

Our concern with  Witte {\it et al.} arose
from their Letter~\cite{WittePRL17} where  the isochoric conductivity
 $\sigma_{\rm ic}$ of aluminum at 2.7 g/cm$^3$ and temperature $T$=0.3 eV
 is calculated using the Kubo-Greenwood (KG) formula via
 density-functional  theory (DFT) and molecular-dynamics (MD)
 simulations. The conductivity obtained with the semi-empirical 
Heyd-Scuseria-Ernzerhof
 (HSE)~\cite{HSE03}  exchange-correlation (XC) functional was
 2.22$\times 10^6$ S/m (see Fig. 1, Ref.~\cite{WittePRL17}, and 
Fig.~\ref{replyXC.fig}), while the Purdew-Burke-Ernzerhof
 (PBE)~\cite{PBE96} functional gave
 3.35 $\times 10^6$ S/m.  Such a a large change (factor of $\sim$ 1.5)
 on changing   XC-functionals for a static property was surprising.  
 The claim of  `better' results in their Fig.1
 using the HSE functional depended
 on just one
 experimental static conductivity  from
 Gathers~\cite{IJTher83} Table-II, column 4., i.e., for
$T$ = 0.3 eV, incorrectly taken to be for {\it isochoric} aluminum.
 In May 2017 this seemed a simple error in reading  Gathers' data
 and we  suggested  a minor correction.

\begin{figure}[t]
\includegraphics[width=\columnwidth]{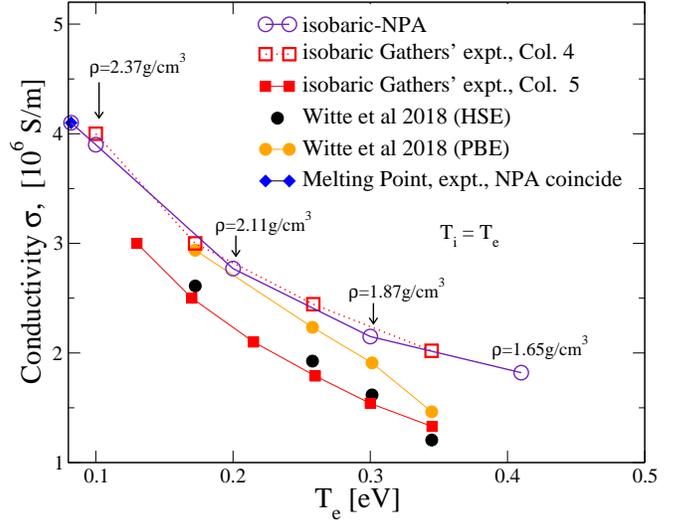}
\caption{(Color online) The isobaric conductivities $\sigma_{\rm ib}$ of
 aluminum from
 its melting point to about 0.4 eV. Here we use Fig. 9 of Witte {\it et al}
revised calculations~\cite{Witte-PhyPls18},  Gather's experimental
data, and our NPA-Ziman calculations~\cite{cond3-17}. We claim that the
 valid experimental
data set is Gathers' Column 4 (empty red squares) which aligns correctly
 with the value at the melting point, while Witte {\it et al} claim that the 
correct experimental data set is Gathers' Column 5 (filled red squares).
In Ref.~\cite{WittePRL17} Witte claimed that Gathers' column 4 data were
 isochoric  conductivities, provoking our disagreement.
}
\label{replyXC2.fig}
\end{figure}

In response, and as justification, Witte {\it et al.}  sent  their
Li-$\sigma$  comparisons with  experimental data~\cite{Fortov02, Bastea02}. 
Our NPA results on Li  agree with theirs, except  at 
 $\rho=0.6$ g/cm$^3$ and  $T=4.5$ eV~\cite{Witte17}.
We  also drew attention to the last row of Table 23 of Gathers'
1986  review article~\cite{Gathers86} where the conductivities
were clearly for the {\it isobaric} densities  of aluminum.

We are satisfied that  Witte {\it et al.} 
have  modified their value for $\sigma_{\rm ic}$ using
the HSC functional, differing from the
$\sigma_{\rm ic}$ from  PBE  by $< 20$\% (Fig.~\ref{replyXC.fig}).
Furthermore, in Ref.~\onlinecite{Witte-PhyPls18} they say:  
 ``Note that the conductivity values reported
 here at $T$= 0.3 eV for
the HSE functional are slightly higher than those published
earlier in Ref. 10, where a part of the non-local contributions
in the transition matrix elements was not taken into account''.
In Ref.~\onlinecite{WittePRL17}
 the conductivity $\sigma_{\rm ic}$ at 0.3 eV is given as 2.22 $\times 10^6$ S/m,
 for 2.7 g/cm$^3$  while their corrected values are 2.82 $\times 10^6$ S/m,
while the value, i.e.,  $\sigma_{\rm ib}$  at 1.87 g/cm$^3$ is
 1.62 $\times 10^6$ S/m.

They now treat the Gathers' experimental  results  as being  isobaric,
thus correcting  our main concern. The difference between column 4 and
column 5 of Gathers is of the order of 10-25\% as seen in Fig.~\ref{replyXC2.fig},
or from Table I of Ref.~\cite{cond3-17}.
Gathers does not measure the temperature, but estimates it from the
heat input. As seen in the discussion in Gathers (even in regard
to his experiments on copper), the raw resistivities, measured as a function
of the input enthalpy $H$ has to be corrected for volume expansion to
determine the temperature associated with that input enthalpy. This
correction is included in Gathers' column 5.

\section{Detailed considerations}
We discuss the issues raised in WitteC in more detail below.

\subsection{Use of the Ziman formula --- Comment item (i)}
The WitteC criticizes the spherically-averaged $S(k)$  approximation
proposed in Sec. II-C of DKHL. The spherical average was used in
calculating ultra-fast conductivities $\sigma_{\rm uf}$ for
two-temperature ($2T$) experiments on polycrystalline samples of
cubic crystals~\cite{Sper15}. The spherical
average applies to polycrystalline materials when the probe beam
samples a large enough volume.

There was no mention of the use of a single crystal in the
LCLS experiment. In fact, a 50 $\mu$m thick
aluminum foil and a beam of up to 10 $\mu$m had been used.
 If Neumeyer's data (cited
as a private communication in the comment by Witte et al) was for
 the same foil and averaging
over the same volume as the LCLS beam, then the report on the
experiment given in Ref.~\cite{Sper15}
needs a correction. When the papers DW16 and DKHL
were written, Ref.~\cite{Sper15} was the only published source of information
on the experiment and there the aluminum was claimed to be a plasma
with $T_i=T_e=6 eV$. Their claim that  $T_i=6$ eV was not
credible, and we explored a 2$T$ model with cold ions but
there was no reason to use a crystal-like model. In Witte's comment they
now claim that the $S(k)$ has the crystalline features of a
cold-ion subsystem. Thus the analysis by Sperling {\it et al}~\cite{Sper15} needs
an erratum to ensure that other readers will be aware of the nature of the
sample used.

An averages $S(k)$ arises naturally with polycrystalline samples.
The laser beam averages over a volume of 10 $\mu$m diameter and 
depth of 50$\mu$m.
The Laue peaks broaden and fill the $k$-regions, as
in the spherical average. The possibility of improving the
theory and experiment using single crystals was already
stated by us in DKHL thus:\\
``{\it  Hence this calculation appears to need further
 improvement for $T < 1.0$ eV, e.g., using the structure factor of
 the FCC solid and ... band-structure effects}''.\\
So WitteC's concern is already voiced by us. No crystalline $S(k)$ was
reported in Ref.~\cite{Sper15} which claimed that the ions were at 6 eV.  
Our papers were based on the available information.

Furthermore Witte {\it et al.}  seem to suggest that the
 NPA method cannot generate the needed solid-state structure factor
 for aluminum at low temperatures.
 This is incorrect. The NPA method applies seamlessly from solid to
liquid to plasma as long as there are `free' electrons in the system.
We even predicted the crystalline $S(k)$ and  calculated
phonons along all standard directions in the
 Brillouin zone~\cite{HarbEOSPhn17}. 
Furthermore, the first applications of the NPA methods were to
 solids at $T=0$ published decades
ago by Dagens~\cite{Dagens75} who calculated ground state energies
and band structures for aluminum and other {\it simple metals}.
Here and elsewhere we use  the appellation `simple metal' in a standard way,
as  used  in Ashcroft and Mermin~\cite{AshMer76} and in other other
standard works. For example, Rossiter (Ref.~\cite{Rossiter87}, p13)
 defines the term  `simple metal' to mean `virtually all electrons at the
 Fermi surface are $s$-like'. 

The NPA results for solids are in good agreement with
 other available methods~\cite{Dagens75}. Our NPA
model is not only a finite-$T$ generalization of earlier models
like those of Dagens,
 but also includes the property
that the NPA is not just an approximation limited to non-overlapping
muffin-tin average atoms. It is  {\it  an exact reduction}
of the $N$-ion DFT problem to a single ion problem when a proper
ion-ion correlation functional is used~\cite{DWP1}. So
we do not use the muffin-tin approximation for the continuum electrons,
commonly invoked in average-atom(AA) models.
 Very little attention to this comprehensive DFT approach has
 been paid, some exceptions being the work of
Chihara~\cite{Chihara86,Chihara16}, Ichimaru {\it et al}~\cite{YanIchimaru91},
 Xu and Hansen~\cite{XuHanson02}.

Only the $S(k)$ near $2k_F$ is relevant to $\sigma$, and  not   $k$ values
outside the window of integration $f(k)\{1-f(k)\}$ in the Ziman
formula. Since Sperling {\it et al.}  used  $S(k)=1$, with not even a
scalar-$k$ dependence, we  proposed
a well-known approximation with about 20\% error for single crystals
and no error for powders. Such approximations, and the nearly-free
 electron  Ziman formula are widely used ~\cite{Rossiter87}.

Witte {\it et al.}  have  {\it not} presented a
DFT-MD-KG calculation of the $2T$-WDM conductivity to  estimate 
(the non-spherical or)  any contribution whatever from their $S(k)$.
In fact, DFT-MD-KG cannot provide  Kubo-Greenwood conductivity
 results for solids or  cold ions
for reasons explained in sec.~\ref{coldIons-DFT.sec}.
the lowest-$T$  $\sigma$ reported by Witte {\it et al.}
 is for  equilibrium   $T_i=T_e=0.15$ eV.


\subsection{Why the DFT-MD-KG conductivity calculation fails for cold ions}
\label{coldIons-DFT.sec}
The lowest temperature where Witte {\it et al.} have reported a conductivity
for aluminum using the DFT-MD-KG method is 0.15 eV with the ion and electron
 temperatures equal, viz.,
$T_i=T_e$. In 2$T$-ultrafast applications $T_i$ is less than the melting
point $T_m$ and the DFT-MD-KG method becomes impractical for such cold ions
as may be clear from the following discussion.

 The DFT-MD-KG method attempts to represent the plasma ions by an ordered
 periodic crystal with a
unit cell containing $N\sim 100-200$  atoms per ionic species for which
 a band structure
is calculated for that particular crystal, although no such bands exist
in a plasma. This highly unrealistic model has to  capture the atomic
and electronic disorder found in the real sample by evolving it in time
 at the temperature
$T_i$ and generating  many realizations of crystal configurations, with
each yielding a dynamic conductivity for the resultant ionic configuration.
Then  a Drude model has to be applied to obtain a
 static conductivity, using
the `mean-free path' approximation for each case, as may be seen
in a typical calculation like Ref. \cite{Witte-PhyPls18}.
Then an  average over all such individual `runs'  is made to obtain a static
conductivity  which  is identified with that of the plasma. This process is
an extremely computer-intensive and also time consuming process, and  becomes
impractical for cold ions which require  many many time steps to generate
new equilibrium ionic configurations. Alternatively, very large unit cells as
recommended by Pozzo {\it et al.}~\cite{Pozzo11} may be required to assure
self averaging.
 
This serious bottleneck probably explains why Witte {\it et al.} have not
provided UF-conductivities  or  equilibrium conductivities (with or without
a spherical average) at the melting point $T_m$.

\subsection{The excellent accord between our XRTs calculation and that
 of Witte et al.\ --- Comment item (ii)}
This comment has two parts.
(a) After displaying the $S(\vec{k})$ of lattice-like aluminum in
 Fig.\ 1 of  the Comment, Witte {\it et al.} state the following.

`` Furthermore, DWD [2] claim, 
`The excellent accord
between our XRTS calculation and that of Witte et al. (sic)
are fully consistent with ...
DFT-MD simulations.' This statement is invalid ... In addition to the
 missing diffraction peaks ...''

Our claim of an excellent accord, given in DKHL Appendix, Sec.1 (lines
690-696, and 705-709) entitled:\\
``1. X-ray Thomson scattering calculation for Li ...''\\
is unambiguously  {\it not} about  aluminum crystals.
No Laue peaks are expected for Li WDM at $T_i=T_e$=4.5 eV.
Fig.\ 9 of Ref.~\cite{cond3-17} displays the
excellent agreement between the DFT-MD and NPA calculations,
irrespective of the XC-functional used.

(b)  WitteC  lists  the
shortcomings of the Ziman formula, already discussed
 by Dharma-wardana {\it et al.} \ in e.g., Refs.~\cite{PDW-Res87,Thermophys99,
cargese94}. Our objective in DKHL was to use
\begin{itemize}
\item{the simplest NPA, XC-functional in the
  local-density approximation (LDA),}  
\item{weak local pseudopotentials,}
\item{the nearly-free electron independent-scatterer Ziman formula.}
\end{itemize}
to assess the quality of the so obtained  WDM conductivities. 
We are well within experimental
error or within 20\%  where  accurate experimental data are available.
large-$N$  DFT-MD-KG, fancy  XC-functionals etc.,
seem to do worse with $\sigma_{\rm ic}$ falling below experimental
$\sigma_{\rm ib}$ (see below, section \ref{Gather-Cor.sec}) for low $T$.
As for average atom (AA) models, they fail for 
low-$T$  where test data are reliable. 

WitteC ignores  the short-comings of  the DFT-MD-KG approach and
provides evidence for the numerical convergence of their simulations
with $N$ up to 216 in their Fig. 4.
As shown in sec.~\ref{Gather-Cor.sec}
  such converged results for the HSE
or PBE functionals significantly violate  known bounds, predicting
 $\sigma_{\rm ic}<\sigma_{\rm ib}$ at low $T$. This does not seem
to be a problem linked to the XC-functional.
MD-KG treats a plasma as an average over a sequence of  crystals
with  band-structure. Only the dynamic conductivity $\sigma(\omega)$  is given
 by the KG formula. The latter uses various
assumptions (e.g., single electron states,  mean free-paths etc., that are also
used in the Ziman formula), as well as the Drude model. Also, 
no valid derivation of the $T_i\ne T_e$ UFM KG formula exits, and the limit
$\omega \to 0$ does not exist. 

The  DFT-MD-KG community uses the Kohn-Sham eigenstates
of fictitious non-interacting Kohn-Sham electrons, instead of 
true electron eigenstates or approximate Dyson eigenstates~\cite{PDW-levelWidth},
(see: DKHL, Sec. 2 of the Appendix, line 816).
Local-field factors (LFFs) $G_{ei}(q)$ that moderate the
electron-ion interaction $U_{ei}(q)$, being dominated by the small-$q$ limit
(c.f.,  compressibility sum rule) are poorly captured by small-$N$
simulations, e.g., $N=64$. Witte et al have given $N=216$ simulations for
some cases and argued that the $q \to 0$ issues cannot be a problem,
 but their  calculations,
while revealing the problem, do not resolve it.  We take up this discussion
further in sec.~\ref{Gather-Cor.sec}.

\section{Witte comment items iii and iv}
These items refer to DW16 and will not be treated here.

\section{Static conductivity of aluminum and average-atom (AA) models ---
 Comment item (v)}
 WitteC's claim that NPA  cannot capture the non-Drude
 behaviour of $\sigma(\omega)$ near $\omega\sim E_F$  is  incorrect.
The NPA   phase shifts  are  used to calculate
 the modified density of states (DOS), impose the Friedel sumrule,
and  calculate a $\bar{Z}$ consistent with the modified DOS (see Eq. A2,
Appendix to DKHL). The non-Drude features arise from the modified DOS.
Such NPA calculations for the DOS agree  with DFT calculations, even for
complex fluids like carbon and Si, as  shown
by Dharma-wardana and Perrot  in 1990~\cite{DWP-carb90}. An effective mass
$m^*$  accounts for the modified DOS in  `simple' metallic
 C, Si systems~\cite{cdw-Utah12}.
A detailed response regarding this is more appropriately relegated to
a proposed reply to the comments by Witte {\it et al} on  DW16.

Figure 3(a) of the Witte comment  shows  DFT $\sigma_{\rm ic}$ for Al
 at $T_i=T_e$  taken to high temperatures $T>E_F$ 
(see Fig.~\ref{mix.fig}). But serious questions arise.\\
(i) WitteC compares equilibrium data with the {\it non-equilibrium}
Milschberg data for $\sigma_{\rm uf}$. This pioneering
experiment~\cite{Milsch88} is irrelevant to $T_i=T_e$
systems. Furthermore, Ref.~\cite{Milsch88}  used an over-simplified
 data reduction,  
and better  models give a different $T$-dependence~\cite{forsman98,Ng05}.
Nevertheless, intuitive plasma models have been constructed to
give conductivity results that mimic the results of the Milschberg data.
Some of these AA models which use the thermodynamic
potential of mean force rather than the electron-ion potential
will be reviewed below, in sec.~\ref{model-calc}. 

(ii) As laser excitations couple with the free electrons,
three electrons per ion ($\bar{Z}$=3) are heated in
Milschberg-type experiments using  initially solid Al.
The $T_i\ne T_e$ Ziman calculations reproduce the 
minimum seen in the conductivity.
Dharma-wardana and Perrot demonstrated such a minimum in 1992
 in Ref.~\onlinecite{CDWP-milsch92} when the ionization
 $\bar{Z}$ is held fixed. However, $\bar{Z}$ 
increases  with increasing $T$.  
Ionic mixtures  with $\bar{Z}=3\to4\to5\to6$ occur for
 $T$ in the 8-50 eV range. DFT-MD  beyond 8 eV, to 15 eV
 (as in the Witte comment) encounters  $Z=3, 4$ ionized Al. Such DFT-MD
simulations need at least some 100 atoms {\it per  species}
in DFT-MD simulations
to obtain  $S(q), G_{ei}(q)$,
and transport properties. Witte {\it et al.} have used
only 64 atoms. This is totally inadequate for ionic mixtures containing
several types of ions with different ionizations.
To clarify  matters, we study such a mixture of ionic species~\cite{eos95}.

\section{Results for a mixture of ionic species.} 
\label{model-calc}
In the NPA model, the free-electron  pileup $\Delta n(r)$
around an aluminum nucleus immersed in the appropriate plasma
medium is used to construct a weak {\it local} pseudopotential whose
Fourier transform, $U(q)$ is given by
\begin{equation}
U(q)=\Delta n(q)/\chi(q)
\end{equation}
This scheme is appropriate for `simple metals' but not transition metals
where the electrons at the Fermi surface are not $s$-like.
Here $\chi(k)$ is the interacting density-density linear-response
function at finite temperature and at the given electron density.
This response function includes finite-$T$ local field factors
constructed to satisfy the compressibility sumrule, as discussed
in previous publications~\cite{cond3-17,eos95}. The resulting
pseudopotentials are fitted to the parametric form given in Eq. 60
of Dhrama-wardana and Perrot~\cite{ELR98}  where the Heine-Abarankov form
is  generalized via the modulating function $M(q)$. This
better accounts for the $q>2k_F$ regime by having two extra parameters
$\lambda$ and $q_0$. The $e$-$i$ interaction
$U_{ei}(q)$ is written here with the subscripts $ei$ suppressed
for brevity.
\begin{eqnarray}
U(q)&=& U_{\rm HA}(q)M(q),\\
M(q)&=&\{1+\lambda(q/q_0)^2\}/\{1+(q/q_0)^2\}.
\end{eqnarray}

\begin{figure}
\includegraphics[width=\columnwidth]{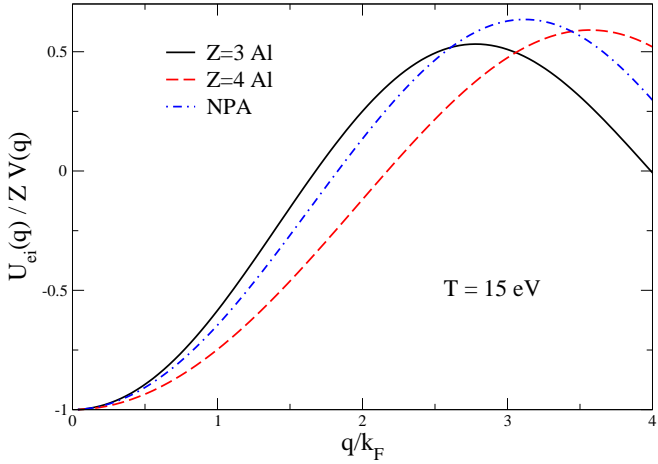}
\caption{(Color online) The pseudopotentials for the  ionization
species $s$=1,2,  with integer ionizations $Z_s$ =3 and 4 are
compared with the NPA pseudopotential with with a mean ionization
$\bar{Z}=3.17$ applicable at $T=15$ eV. The quantity plotted is
 the pseudopotential $U_{es}(q)/ZV_q$, i.e.,
in units of the bare electron-ion interaction $ZV_q=4\pi Z/q^2$
(atomic units).The plasma density is 2.7 g/cm$^3$ with a mean ion-ion
$\Gamma \simeq 6.1$.}
\label{pseudo.fig}
\end{figure}

\begin{table}
\caption{The composition fractions $x_s, s=1,2,3$ for the
 ionic species Al$^{z+}$,
with integer ionizations $Z_s=3, 4,  5$  in the temperature range 10-40 eV
for an aluminum plasma at the density $\rho=2.7$ g/cm$^3$.}
\begin{ruledtabular}
\begin{tabular}{cccc}
T [eV]  &  $x_1, Z=3$ & $x_2, Z=4$ & $x_3, Z=5$  \\
\hline
10      &   0.970     & 0.030      & 0.00        \\
15      &   0.830     & 0.170      & 0.00        \\
25      &   0.110     & 0.890      & 0.00        \\
35      &   0.0       & 0.300      & 0.70        \\
38.8    &   0.0       & 0.001      & 0.99        \\
\end{tabular}
\end{ruledtabular}
\end{table}

The pseudopotentials for the individual species are used to
construct the pair-potentials $V_{ss'}(q)$,  given by:
\begin{equation}
V_{ss'}(q)=Z^2V_q+U(q)_{es}(q)\chi(q)U_{es'}(q).
\end{equation}
They  are used in a multi-species
hyper-netted-chain (HNC) calculation yielding the pair-distribution
functions (PDFs) $g_{ss'}(r)$, and structure factors
$S_{ss'}(q)$. These can be used to construct the equation of
state (EOS) of the multi-species plasma in the HNC approximation
as detailed in Ref.~\cite{eos95}. For temperatures below 1 eV,
a bridge-diagram correction is included in the HNC, using the
hard-sphere ansatz as detailed in Ref.~\cite{PeBe}. That is, we
use the modified HNC equation with a
bridge contribution that assures that the compressibility obtained
from the EOS is in agreement with the $S_{ii}(q\to 0)$ limit. The
simple HNC equation is sufficient for $T>$ 1 eV.
Although the appendix to Ref.~\cite{eos95}
presented the details of the
core-core polarization contributions to the pair potential, we neglect
them in this study. Since our focus here is the conductivity, we
do not discuss the EOS results, but consider the
Ziman formula for a mixture of ions giving the resistivity
$R=1/\sigma$.

\begin{eqnarray}
\label{ziman-mix.eq}
R&=&\frac{\hbar}{e^2}\frac{1}{3\pi}\frac{\rho}{\bar{n}^2}\\
 & &\times \int_0^\infty(1+\exp\{\beta q^2/8-\mu\})^{-1}q^3\Sigma(q)dq,\\
\Sigma(q) &=&\sum_{s,s'}(x_sx_s')^{1/2}S_{ss'}(q)
\frac{U_{es}(q)U_{es'}(q)}
{\{2\pi\varepsilon(q)\}^2}  
\end{eqnarray}

\begin{figure}[t]
\includegraphics[width=\columnwidth]{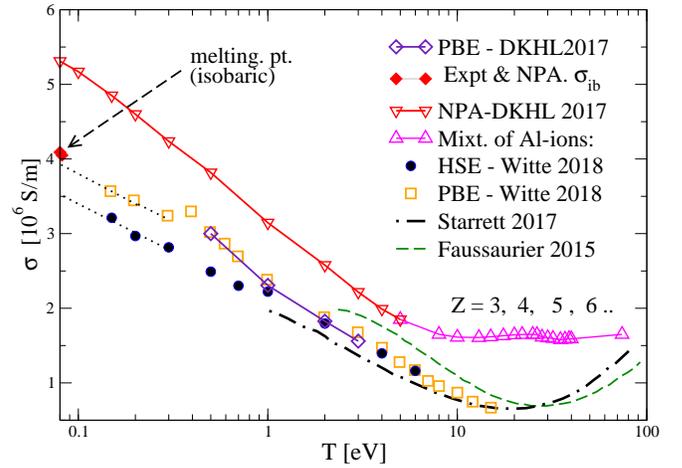}
\caption{(Color online) The  $\sigma_{\rm ic}$ from two recent
AA models  display a broad minimum for 20-30 eV. The 64-atom simulations
of Witte {\it et al.}  show the same trend. For $T > 5$ eV the
conductivity of the  multispecies NPA plasma~\cite{eos95}
shows a weak rise in $\sigma_{\rm ic}$ at 26.34 eV when  $Z=4$ holds.
The MD-DFT-KG $\sigma_{\rm ic}$, $T\to 0.082$ eV  incorrectly
extrapolate to below the red
 diamond, i.e., the {\it isobaric} melting-point conductivity.
 See Eqn.~\ref{predict.eqn} for
 the  consistency of  $\sigma_{\rm ic}\simeq 5\times 10^6$ S/m obtained
 from the NPA  at 0.082 eV with the experimental isobaric value.
 See Sec. VI of SM for details.
}
\label{mix.fig}
\end{figure}

The resulting conductivity for an ionic mixture is displayed
 in Fig.~\ref{mix.fig}.
together with other calculations. It shows that DFT-MD-KG calculations
 give a
significantly lower isochoric conductivity than from the NPA-mixture
 calculation, and
extrapolates to a value below the isobaric conductivity at the melting point.
One possible problem is the limitation of the DFT-MD-KG calculations  to
64 atoms; 
In a 64-atom simulation
 with even two species, .e., 32 atoms/per species, 
the `surface' atoms of the simulation
cluster dominate over
the `bulk-like' atoms. The LFF $G_{ei}(q)$ is known to be mostly
set by the value of $G_{ei}(q\to 0)$ due to the importance of the
compressibility sum rule. But this is poorly captured by small-$N$
simulations in much the same way as how they poorly reproduce
 the small-$q$ behaviour of $S(q)$. From  Fig. 4 of  WitteC we see
that a 32 atom calculation may significantly underestimate a
 larger-$N$ calculation. In effect, if there are $m$ ionic species
in the system, the simulation must use  $\sim100m$ 
particles to have the quality of a 100-atom single species simulation. 
(see also Sec.~\ref{convergence.sec}). 

It should however be noted that the calculations of Witte in the
 low temperature
range (where $Z=\bar{Z}=3$) cannot be faulted for the use of a
 small $N$ since they
have shown results for $N=214$. Hence the under-estimate of the 
low-temperature
$\sigma_{\rm is}$ (and indeed even $\sigma_{\rm ib}$) become an
 intriguing problem. Perhaps
 very large $N$ simulations are needed, as posited by Pozzo {\it et al}
for the case of the low-temperature conductivity  of WDM sodium.

\section{The role  of electron-electron and ion-ion interactions
in electron scattering in the 5 eV to 50 eV regime.}
\label{pdfs.sec}
We examine Starrett's  approach~\cite{Starrett17} to define what he
explains to be the
``interaction potential between the scattering electron and the ion
 so that it correctly includes the effects of ionic structure,
 screening by electrons and partial ionization''. As already stated,
various attempts  to use the potential of mean force and the
XC-potential as  scattering
potentials need a  rigorous derivation based on the
current-current correlation function or some such basic transport
theory.

\begin{figure}
\includegraphics[width=\columnwidth]{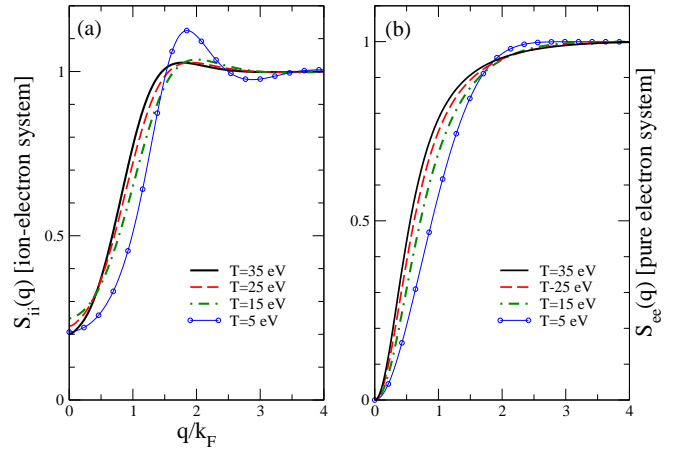}
\caption{(Color online)(a) The ion-ion structure factor $S_{ii}(q)$
from the NPA potentials, for aluminum plasmas (2.7 g/cm$^3$) with $T=5$ eV
 to 35 eV. (b) The electron-electron $S(q)$ for a {\it pure} electron fluid
at the densities and temperatures corresponding to those of the aluminum
plasma is given mainly to illustrate the weak coupling nature of the fluid.}
\label{sk.fig}
\end{figure}

The Ziman formula is based on the force-force correlation
function. The Ziman formula that we use for these calculations, viz.,
Eq.~\ref{ziman-mix.eq} uses a single scattering center and the
effect of the other scatters is brought in via the ion-ion
partial structure factors $S_{ss'}(k)$ which are related by
 a Fourier transform
to the pair correlation functions $h_{ss'}(r)=g_{ss'}(r)-1$. The
single scatterer model fails in many circumstances and a multiple
scattering model is then needed. A measure of the strong
coupling present in a plasma is given by the parameter
 $\Gamma=Z^2/(r_{ws}T)$.
For our system, ($T,\Gamma$) is such that we have  (10,8.4);
(20,5.5); (30,5.6);  (40,5.9); (60,5.9); (80,6.0).
These numbers show the  phenomenon of the approximately
 constant `$\Gamma$-plateau'
 that has been discussed by Cl\'erouin {\it et al.}~\cite{Clerouin14Gamma}.
The Al-plasma in the regime where the AA models
find  a broad conductivity minimum has a $\Gamma\sim 5-6$. The Al-Al
average-atom ion-ion structure factor $S_{ii}(q)$ for the cases $T$ =5, 15,
25, 35 eV are shown in Fig.~\ref{sk.fig}. Hence we believe that
multiple-scattering effects are irrelevant in this regime, and the Ziman
formula, i.e., Eq.~\ref{ziman-mix.eq} holds.

Furthermore, we look at the ion-ion and electron-electron pair-distribution
functions (PDFs) relevant to the aluminum plasma of density 2.7 g/cm$^3$ in
the temperature range 5-50 eV. If phase-shifted plane-wave states are an
adequate representation of the wave functions of the continuum electrons,
then electron-electron scattering does not contribute to the
resistivity as the electron momentum is a `good quantum number'~\cite{resAlAu06}. 
In contrast, the Boltzmann equation uses no concepts of eigenstates,
and {\it assumes} momentum randomization after  each collision and
always contains e-e contributions to electron scattering. In any case,
 e-e scattering calculated within
such models   makes only a minor contribution  to the resistivity in
 high-$Z$ materials like aluminum. Hence it is unlikely that Starrett's
calculations would contain significant e-e scatting contributions to
the conductivity.

Electron-electron scattering can be relevant in systems where the
particulate nature of the electron subsystem becomes relevant.
If the electron distribution
is  `grainy' or `disordered' within the relevant length scale,  then the
 system is in the `diffusive regime' where planewave states are no-longer
 good eigenstates of the Hamiltonian. In such systems the electron-electron
 structure factor and the pair-distribution function will show
the characteristics of a disorder-dominated fluid, in the diffusive regime.
Then indeed e-e  interactions may contribute to the electrical resistivity.
This is surely not the case here.

The XC-potential acts on (fictitious) Kohn-Sham electrons which are
non-interacting but at the density of the interacting system to provide a
many-body correction to the free energy. It does not
act on ``real'' electrons as applicable in conductivity equations.

\begin{table}
\caption{
\label{rnl.tab}
The average bound-state radii $<r_{nl}>$ (in a.u.) of the NPA model
 for  the 2$s$  and 2$p$ atomic states of the aluminum plasma
in the temperature range 10--35 eV, density $\rho=2.7$ g/cm$^3$,
$r_{ws}\simeq 3$ a.u.
}
\begin{ruledtabular}
\begin{tabular}{cccc}
$T$ [eV]  &   $<r_{2s}>$    & $<r_{2p}>$        \\
\hline
10        &   0.984         & 0.016             \\
15        &   0.830         & 0.170             \\
25        &   0.110         & 0.890             \\
35        &   0.5814        & 0.5384            \\
\end{tabular}
\end{ruledtabular}
\end{table}

The ion-ion structure factor $S_{ii}(q)$ and the electron-electron
structure factor $S_{ee}(q)$ are of interest in understanding the
interplay of many-body interactions. The $S_{ii}(q)$ obtained from
the NPA model is readily available and shown in Fig.~\ref{sk.fig}(a).
The agreement of the NPA calculated  $S(q), g(r)$ with those from
X-ray scattering experiments on liquid metals, and with DFT-MD has
 been established in many publications, the most recent being
 in Ref.~\cite{HarbourDSF18} in regard to aluminum.

An approximation to $S(q)$ can be readily calculated in the
 $\Gamma$-plateau
region using the classical one-component plasma (OCP) model containing only
ions neutralized by a uniform static background. Such an approximation
will get the `main peak' region right, but it  will be seriously wrong
 mainly in the small-$q$ region, unlike the result
 obtained via the NPA where the compressibility sumrule is satisfied.
The $S_{ee}(q)$ shown in Fig.~\ref{sk.fig}(b) is  for a fully interacting
pure electron plasma (a quantum OCP) where quantum effects are included via
 the classical-map hyper-netted-chain (CHNC) procedure~\cite{prl1}. We do not
 attempt to calculate the $S_{ee}(q)$ in the presence of the ions as such
 a calculation satisfying the sumrules as well as the quantum mechanics of
 the coupled electron-ion system is a more complex task not required for
 this study, requiring the use of a three component CHNC model for aluminum.

However, the main change will be the
 modification of the small-$q$ region of $S_{ee}(q)$ to conform with those
 of $S_{ii}(q)$ so as to satisfy the compressibility sum rule for an
 electron-ion system. However, the essential  point to note is the
clearly  weakly-coupled  nature of the interactions
 present in these systems, as well as the fact that the electron-ion
 interaction  (Fig.~\ref{pseudo.fig}) can be given as a  weak local
 pseudopotential.
 Hence, in our view, the attempt to
 invoke e-e scattering and contributions to the scattering potential
 from ion-ion interactions, other than those which  enter in the standard
 theory  via the LFF $G_{ei}(q)$ of e-i interaction, and the LFFs that
 enter into  the dielectric function, remains unsubstantiated even
 within an intuitive  physical picture.

\begin{figure}[t]
\subfigure{\label{fig:edge-a}\includegraphics[width=0.36\columnwidth,height=4.9cm]{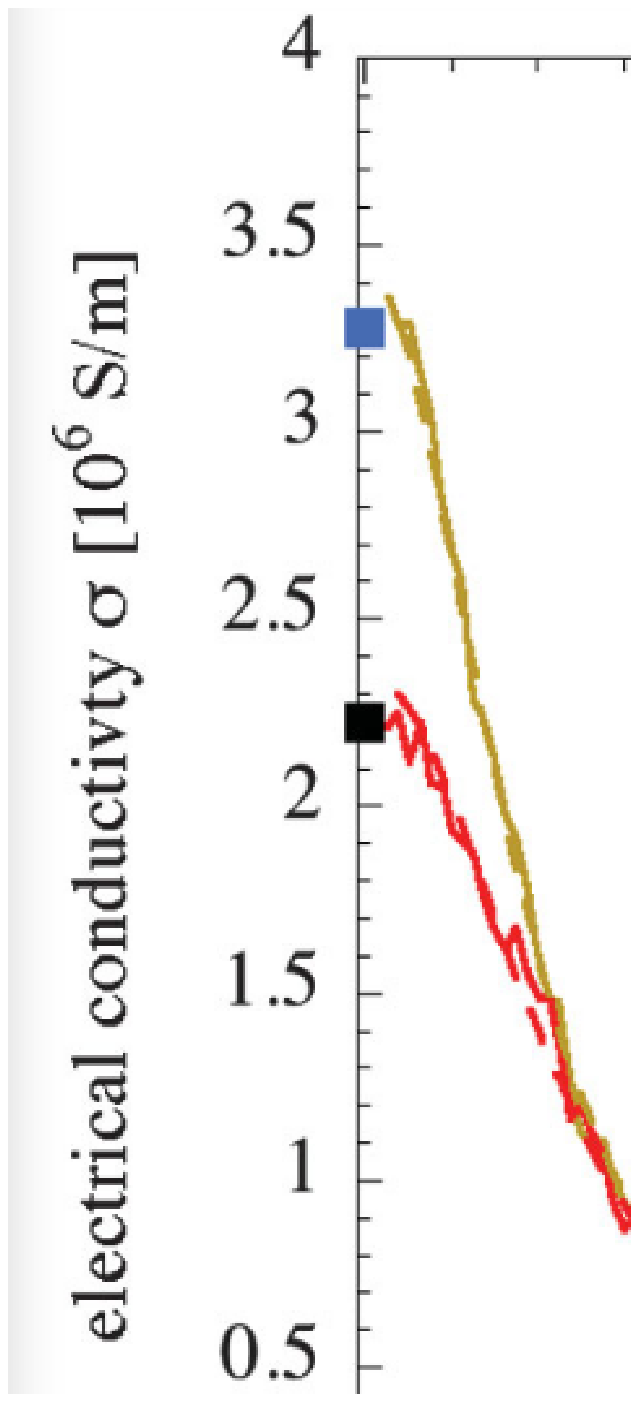}}
\subfigure{\label{fig:edge-b}\includegraphics[width=0.59\columnwidth,height=4.8cm]{sigmaSml.eps}}
\caption{(Color online) {\bf Left Panel}: Extract from Fig. 1 of  Witte {\it et al.}~\cite{WittePRL17}
displaying converged $\sigma_{ic}(\omega)$ plots for isochoric-Al at 0.3 eV for
 small $\omega$.  The upper curve  extrapolating towards 3.5 is using the PBE, while
 the lower curve  (extrapolating towards $\sim$ 2.4) is obtained using the HSE functional.
{\bf Right Panel}:  The behaviour of  $\sigma_{ic}$ as $T\to T_m$ from different
calculations. These presumably ``well-converged'' DFT-MD-KG results
 extrapolate to a value  even {\it below} the known experimental isobaric
 conductivity of aluminum at its melting point $T_m=0.082$ eV~\cite{Leavens81}.}
\label{sigmaSml.fig}
\end{figure}

\section{Kohn-Sham level structure of the aluminum plasma}
\label{KS-levels.sec}
Average atom calculations as well as NPA calculations require
special procedures if bound-states stretch out beyond the Wigner-Seitz
sphere of the ion, as is typically the case for transition-metal  ions.
The Wigner-Seitz radius $r_{ws}\simeq 3$ a.u. for aluminum at
2.7 g/cm$^3$. In Table~\ref{rnl.tab} we present the radial extension of
 the $n=2$ bound states of  the NPA  for
 typical cases in the temperature range  $T$=10 to 35 eV.
The NPA bound states are compactly within the Wigner-Seitz sphere and no
difficulty arises from bound-like continuum resonances, and
`discontinuous' changes in $\bar{Z}$ in the NPA. The
bound levels are compact in the respective ionization states
(Z=3,4, 5 models) needed for the multi-species plasma.
\section{convergence of DFT-MD-KG calculations}
\label{convergence.sec}

In  WitteC it is stated that
`` DWD finds larger dc conductivities from NPA-Ziman
calculations compared to DFT and attributes this partly
to the `inability of the DFT-MD-KG approach to access
small-$k$ scattering contributions unless the number
of atoms $N$ in the simulation is sufficiently large.' The
criticism on Refs. [4, 36], however, is not valid. Note that
several earlier studies reported well-converged conductivity
 calculations for aluminum and lithium with similar
particle numbers [35, 37-39]''.

The issue is not just numerical convergence as a function of $N$
as seen by the
increasing closeness of the $\sigma(\omega)$  curves, but also the
question of whether the results approach the expected experimental values,
or converge to a different bound.
In the right panel of Fig.~\ref{sigmaSml.fig} we show results from
two presumably well-converged calculations for $\sigma_{ic}$ of
aluminum (2.7 g/cm$^3$) given by Witte {\it et al.}, one of which uses the
first-principles PBE XC-functional, while the lower estimate of $\sigma_{ic}$
uses the semi-empirical HSE functional which incorporates an {\it ad hoc}
Hartree-Fock contribution adjusted to get bandgaps correct.

We have tentatively indicated an
extrapolation towards the melting regime (0.082 eV for isobaric Al)
using thin dotted lines, for these two sets of calculations. These show
 that {\it both} KG calculations
predict a $\sigma_{ic}$ at 0.082 eV that are {\it even lower} than the
experimentally known isobaric conductivity of 4.1 $\times 10^6$ S/m.
In fact, the most recent `well-converged' HSE
 prediction~\cite{Witte-PhyPls18}
extrapolates towards 3.5 $\times 10^6$ S/m. The isochoric
conductivity of aluminum, densty 2.7 g/cm$^3$  {\it cannot}
  be any smaller than the  isobaric conductivity at the lower  density
of 2.375 g/cm$^3$, 
but the KG simulations strongly contravene this.

The  scaling of the known experimental isobaric conductivity to the
 isochoric  conductivity can be discussed as follows, where we correct
an expression given by Witte et al. in their comment.

Witte {\it et al.} relate $\sigma_{\rm ib}$ to $\sigma_{\rm ic}$
incorrectly using only the volume change. They
ignore the change in the scattering cross-section
 $\Sigma$ due to the changed screening. But in bringing out this
discussion,  Witte {\it et al} seem to recognize the
{\it physical requirement} resulting from the consequent physics.
That is:
\begin{equation}
\label{physical.eqn}
\sigma_{\rm ic}>\sigma_{\rm ib}\; \mbox{for Al}. 
\end{equation}
At low $T\ll E_F$, scattering
 occurs with a momentum
 transfer of $\sim 2k_F$ where $\Sigma$
$\sim |V(2k_F,\rho)|^2$, with $\Sigma\propto
\{\bar{Z}/(|4k_F^2+k_{sc}^2|)\}^2$. Here $k_{sc}$ is
 the $T,\rho$ dependent Thomas-Fermi screening wavevector.
Thus,
 \begin{eqnarray}
\label{predict.eqn}
 \sigma_{\rm ic}&=&\sigma_{ib}\frac{\rho_{\rm ic}}{\rho_{ib}}|X_{\rm ic}/X_{\rm ib}|^2 \\
X_{\rm ib}&=& |4k_F^2+k_{sc}^2|_{{\rm at}\;\rho_{\rm ib}};\;\;\mbox{similarly, for}\;
 X_{\rm ic}.
 \end{eqnarray}
 This provides a consistency test (at very small $T/E_F$)
 relating  the experimental $\sigma_{\rm ib}$ and  $\sigma_{\rm ic}$.

This equation shows that
 $\sigma_{ic}$ should be $\sim$ 5  $\times 10^6$ S/m at the melting point. While the NPA
 correctly captures this value, the KG simulations fall far short.
 In fact, the the HSE functional which
is strongly promoted in Witte {\it et al.}~\cite{WittePRL17}
 gives a $\sigma_{ic}$ that is
 a gross  underestimate of the true $\sigma_{ic}$ by 150\%. In
DKHL we merely gave some suggestions as to why such a discrepancy
 should arise.
These were (i) theoretical short-comings of the KG formula
(ii) possible limitations in the $N$-simulation to
accurately capture the LFFs that are implicit in the effective
electron-ion scattering which are dominated by th $q\to 0$ limit.
However, given the extensive simulations of Witte et al, item (ii)
may not be relevant. However,
the $N$-convergence studies up to $N$=216 presented in  WitteC
 provide no  solution to the puzzle.

 All DFT XC functionals should
 recover the STATIC properties of simple or  complicated
 material systems, but they {\it need not} recover band gaps and
 spectra as DFT is NOT a theory of such properties. So, if
HSE gives better bandgaps and features of the excitation
spectrum,  Cooper minima etc.,  due to the inclusion of {\it ad hoc}
corrections into it, those seeming improvements cannot be
at the sacrifice of {\it static} properties. If static properties
related to the total free energy or the ground state are
sacrificed by a proposed XC functional, then it has fallen
outside DFT variational principles.

It is clear that having seemingly $N$-converged KG simulations
do not guarantee an accurate prediction of the conductivity. Thus
fig. 4(c), (d) of  WitteC merely establish that their best
converged results actually
strongly violate the physical condition given in Eq.~1.
In regard to Fig. 4 (d) of WitteC, what one would like to know is the estimate of
the compressibility from the $S(q\to 0)$ limit at each $N$ and the
extent of its disagreement with the compressibility obtained from the
corresponding equation of state, for PBE and HSE.

Interestingly, the offset between the expected value of $\sigma_{ic}$
at $T_m=0.82$ eV, viz., 5.1 $\times 10^6$ S/m and the PBE or
HSE-$\sigma_{ic}=3.4\times 10^6$ S/m is nearly the same as the offset of the
broad minima of the NPA mixture calculation with those from DFT-MD-KG and
AA calculations in the 25 eV range.

\subsection{Gather's corrections of raw experimental data - 
Witte comment vi}
\label{Gather-Cor.sec} 
We are satisfied that Witte {\it at al.} now
 recognize that Gathers' data are isobaric. Gathers' column 4,
 column 5 data  differ  only by $1bout 0-25$\%, and the difference
 arising from XC-functionals is similar. Hence our main concerns
are resolved, irrespective of the data column used.
 Nevertheless, we examine WitteC, item (vi) further.

Gathers {\it explicitly states} that the  data are applicable
only in the density range $2.42 \ge \rho \ge 1.77$ g/cm$^3$
in both Ref.~\onlinecite{Gathers86} and in  Ref.~\onlinecite{IJTher83}.
He states that the enthalpy $H$ needs a dilation correction $v/v_0$
and this not a crude correction to get $\sigma_{\rm ic}$  
at 2.7 g/c$m^3$. 
And yet, even though extrapolation of the conductivity data
to 2.7 g/cm$^3$ is explicitly forbidden, Witte {\it et al.}  say
 ``{\it We agree that
this extrapolation to $\sigma_{\rm ic}$  made by Gathers is crude ...}''. 

Gathers' 1986 review~\cite{Gathers86} was after
Desai's work~\cite{Desai84}. The converged HSE $\sigma$ of WitteC falls
below Gathers' data~\cite{IJTher83,Gathers86} by  $\sim$20\%.

\subsection{Comment on the XC-functional --- item (viii) of Comment}
 The {\it static} conductivity is an
 equilibrium ensemble property and should be
 captured by  standard DFT. Given that Witte {\it et al.} have revised
their value of $\sigma_{\rm ic}$ from the HSE functional, and since even
AA models can pick up the non-Drude behaviour of $\sigma(\omega)$ as stated
by WitteC, there is finally no compelling evidence in favour of
 the  HSE functional which has an {\it ad hoc} inclusion of a quarter of
a Hartree-Fock term.

\section{Conclusion}
The Witte-comment  and also 
 Ref.~\cite{Witte-PhyPls18} show a welcome revision
where they no longer claim that the static conductivity of aluminum at
 $T=0.3$ eV and density 2.7 g/cm$^3$ varies by large factors like
  $\sim$ 1.5 on changing the  XC-functional from PBE to HSE. The
{\it ad hoc} HSC functional gives less  satisfactory static $\sigma$
predictions. 
 Equilibrium isochoric aluminum conductivity
 at higher $T$~\cite{Witte-PhyPls18} where there are several ionization states
will require simulations with some 200-300 atoms for credible results.

\end{document}